%% file: main.tex
\documentclass[sigplan,10pt]{acmart}
\renewcommand\footnotetextcopyrightpermission[1]{}
\AtBeginDocument{%
  }
    
\usepackage{enumitem}
\usepackage{algorithmic}
\usepackage[ruled,linesnumbered, vlined]{algorithm2e}
\usepackage{ulem}
\usepackage{multirow}
\usepackage{pifont}
\usepackage{tikz}
\usepackage{bm}
\usepackage{balance}
\newcommand*\circled[1]{\tikz[baseline=(char.base)]{
            \node[shape=circle,draw,white, fill=black,inner sep=1pt] (char) {#1};}}

\newcommand*\squared[1]{\tikz[baseline=(char.base)]{
            \node[shape=rectangle,draw,white, fill=black, minimum size=10pt, inner sep=1pt] (char) {#1};}}
\newcommand{\nonl}{\renewcommand{\nl}{\let\nl\oldnl}}

\setcopyright{none}

\startPage{1}

\begin{document}


\title{EcoServe: Enabling Cost-effective LLM Serving with Proactive Intra- and Inter-Instance Orchestration}




\author{Jiangsu Du}
\affiliation{%
 \institution{Sun Yat-sen University}
 \city{Guangzhou}
 \country{China}}
\email{dujiangsu@mail.sysu.edu.cn}

\author{Hongbin Zhang}
\affiliation{%
 \institution{Sun Yat-sen University}
 \city{Guangzhou}
 \country{China}}
\email{zhanghb55@mail2.sysu.edu.cn} 

\author{Taosheng Wei}
\affiliation{%
 \institution{Sun Yat-sen University}
 \city{Guangzhou}
 \country{China}}
\email{weitsh@mail2.sysu.edu.cn} 

\author{Zhenyi Zheng}
\affiliation{%
 \institution{Sun Yat-sen University}
 \city{Guangzhou}
 \country{China}}
\email{zhengzhy37@mail2.sysu.edu.cn} 

\author{Kaiyi Wu}
\affiliation{%
 \institution{Sun Yat-sen University}
 \city{Guangzhou}
 \country{China}}
\email{wuky33@mail.sysu.edu.cn} 

\author{Zhiguang Chen}
\affiliation{%
 \institution{Sun Yat-sen University}
 \city{Guangzhou}
 \country{China}}
\email{chenzhg29@mail.sysu.edu.cn} 

\author{Yutong Lu}
\affiliation{%
 \institution{Sun Yat-sen University}
 \city{Guangzhou}
 \country{China}}
\email{luyutong@mail.sysu.edu.cn} 

\begin{abstract}
Existing LLM serving strategies can be categorized based on whether prefill and decode phases are disaggregated: non-disaggregated (NoDG) or fully disaggregated (FuDG).
However, the NoDG strategy leads to strong prefill-decode interference and the FuDG strategy highly relies on high-performance interconnects, making them less cost-effective. 

We introduce EcoServe, a system that enables cost-effective LLM serving on clusters with commodity interconnects.
EcoServe is built on the partially disaggregated (PaDG) strategy, applying temporal disaggregation and rolling activation for proactive intra- and inter-instance scheduling.
It first disaggregates the prefill and decode phases along the time dimension within a single instance to mitigate inter-phase interference and enhance throughput.
Next, it coordinates multiple instances and cyclically activates them to ensure the continuous availability of prefill processing, thereby improving latency.
Thus, EcoServe's basic serving unit is the macro instance, within which multiple instances collaborate.
It further integrates an adaptive scheduling algorithm to route requests in a macro instance and a mitosis scaling approach to enable fine-grained capacity scaling.
Beyond delivering high goodput, EcoServe excels in load balancing, hardware cost, parallelism compatibility, and even engineering simplicity compared to existing solutions.

When serving 30B- and 70B-scale models on a production-level cluster with 32 NVIDIA L20 GPUs using commodity Ethernet, EcoServe averagely improves goodput by 82.49\%, 86.17\%, 122.76\%, and 126.96\% over four representative NoDG and FuDG systems.


\end{abstract}

\keywords{}

\maketitle

\input{tex/Introduction}
\input{tex/Background}
\input{tex/Design}

\input{tex/Evaluation}
\input{tex/Related}

\input{tex/Discussion}
\input{tex/Conclusion}


\clearpage
\balance
\bibliographystyle{ACM-Reference-Format}
\bibliography{sample-base}


\end{document}

%% file: tex/Introduction.tex
\section{Introduction}

Large language models~\cite{touvron2023llama, grattafiori2024llama, brown2020language} (LLMs), have been widely adopted across various tasks~\cite{chatgpt, Characterai, github2023cop, cursor2025}.
To handle the massive LLM requests, optimizing cost per request while ensuring response times meet service level objectives (SLOs) becomes a primary goal.
LLM inference consists of two distinct phases, the prefill phase and the decode phase, each associated with a different SLO, time to first token (TTFT) for the prefill phase and time per output token (TPOT) for the decode phase.
The interplay between TTFT, TPOT, and throughput forms an inherent performance trade-off triangle, in which improving one often comes at the cost of the others.

Existing cluster-level LLM serving solutions~\cite{distserve, qin2024mooncake, patel2024splitwise, faster2024transformer, vllm2024github, yu2022orca, agrawal2024taming} can be categorized into two strategies based on whether prefill and decode phases are disaggregated: the non-disaggregated (NoDG) strategy and the fully disaggregated (FuDG) strategy.
However, both strategies have limitations, either incurring severe interference between prefill and decode phases, or relying heavily on additional hardware capabilities, which prevents them from achieving cost-effectiveness.

\begin{figure}[!t]
    \centering
    \includegraphics[width=0.98\linewidth]{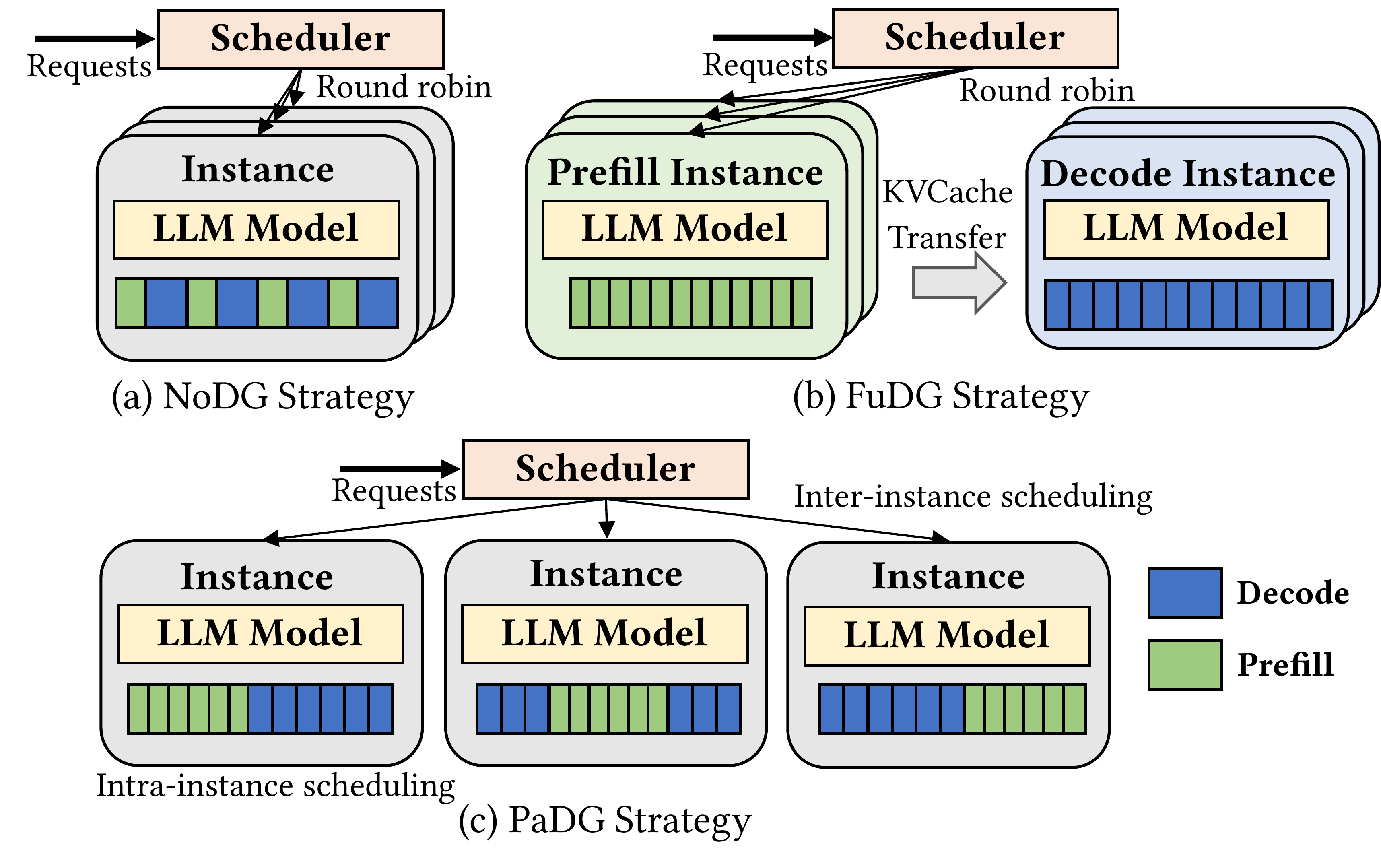}
    \vspace{-10pt}
    \caption{The NoDG, FuDG, and PaDG strategies.}
    \label{fig:twodg}
    \vspace{-10pt}
\end{figure}

Given that the prefill and decode phases share model weights and KV cache, as in Figure~\ref{fig:twodg}(a), the NoDG strategy~\cite{faster2024transformer, vllm2024github, yu2022orca, agrawal2024taming}, which the prefill phase and the decode phase are placed in a single instance, appears to be a natural choice. 
Such colocation inevitably leads to significant interference between the two phases~\cite{distserve}.
For example, if prefills are prioritized and inserted excessively for good TTFT, ongoing decodes are poised to experience longer delays, resulting in poor TPOT.
Prioritizing the scheduling of one phase risks violating the latency requirements of the other, and this interference also harms throughput, as the decode phase cannot accumulate a sufficiently large batch size to saturate GPU resources.
Moreover, the NoDG strategy cannot efficiently adopt pipeline parallelism.
Since prefills are varied in lengths and decodes exhibits tight dependency between iterations, micro batch workloads are generally imbalanced and interdependent, resulting in severe pipeline bubbles and further degrading the NoDG strategy's efficiency.

As illustrated in Figure~\ref{fig:twodg}(b), the FuDG strategy~\cite{distserve, qin2024mooncake, patel2024splitwise} proposes to fully eliminate the prefill-decode interference by assigning the two phases to separate instances.
However, since this strategy requires transferring massive amounts of KV cache between prefill and decode instances, it relies on hyper-clusters with powerful interconnects as the default hardware infrastructure. Unfortunately, high-performance interconnects, such as intra-node NVLINK and inter-node InfiniBand, are not only exceptionally expensive but also power-intensive, even compared to the cost and energy demands of GPUs. Moreover, scaling the performance of FuDG strategy involves both prefill and decode instances, whereas, adjusting the ratio of these two types of instances is challenging, which may lead to significant load imbalance~\cite{liang2025injecting}.
In addition, on nodes without GPU-direct interconnects, tensor parallelism and KV cache migration intensively contend for PCIe bandwidth, potentially becoming a bottleneck.

In this work, we present EcoServe, an LLM serving system designed to deliver cost-effective LLM inference on clusters with commodity interconnects.
As illustrated in Figure~\ref{fig:twodg}(c), our key observation is that intra-instance scheduling, which determines when to execute prefills and decodes, must be coordinated with inter-instance scheduling, which decides when and where requests should be routed, to raise the upper bound of the trade-off triangle and fully utilize available resources.
To this end, our EcoServe is built on the PaDG strategy, which incorporates temporal disaggregation and rolling activation to achieve proactive intra- and inter-instance scheduling.

The PaDG strategy proactively disaggregates the prefill and decode phases along the time dimension within a single instance.
In other words, each instance periodically switches between prefill and decode phases, with each phase lasting longer to reduce switching overhead.
Since both phases are still in a single instance, the PaDG strategy avoids KV cache transmission, unlike the FuDG strategy.
By mitigating the prefill-decode interference, this approach allows EcoServe to achieve significantly higher throughput.

Next, to meet SLOs, the PaDG strategy further employs a rolling activation scheduling.
Without it, if a request is assigned to an instance that has just switched to process the decode phase, it would suffer from an unacceptably high TTFT.
Rolling activation proactively coordinates multiple instances in a cyclic pattern.
At any given time, there are instances specifically activated for processing prefills, capable of delivering acceptable TTFT latency.
A group of such cooperating instances is referred to as a macro instance.
Consequently, EcoServe can theoretically satisfy SLOs while also achieving a higher overall throughput, leading to more cost-effective LLM serving.
Moreover, since the PaDG strategy minimizes prefill-decode switches and KV cache transmission, it is highly compatible with both tensor parallelism and pipeline parallelism.

With the fundamental concept, EcoServe further includes an adaptive scheduling algorithm and the mitosis scaling approach.
The adaptive scheduling algorithm guides request scheduling within the macro instance.
While prioritizing the maintenance of satisfactory TPOT, it identifies the most suitable instance for admitting new requests and determines the optimal number of prefill tokens that can be inserted into that instance.
The mitosis scaling approach enables elastic and fine-grained control over system capacity by continuously adjusting the number of instances within a macro instance, and triggering a split or merge operation when the instance count exceeds predefined thresholds.
To transparently migrate instance between macro instances, it introduces a serializable proxy object that enables logical migration without instance re-initialization and execution interruption .
Our contributions are:
\begin{itemize}[leftmargin=*]
    \item We present EcoServe, a LLM serving system that better enables cost-effective LLM inference on clusters with commodity interconnects.   
    \item We introduce the PaDG strategy, along with the adaptive scheduling algorithm and the mitosis scaling approach. 
    \item We implement EcoServe in a hierarchical architecture.
    \item We evaluate EcoServe and compare it with four representative serving systems, i.e. vLLM, Sarathi, DistServe, and MoonCake.
\end{itemize}




%% file: tex/Background.tex
\section{Preliminary and Motivation}

\subsection{Computation and Memory of LLM Inferences}
\label{sec:computationfeatures}

\begin{figure}[!t]
    \centering
    \includegraphics[width=0.95\linewidth]{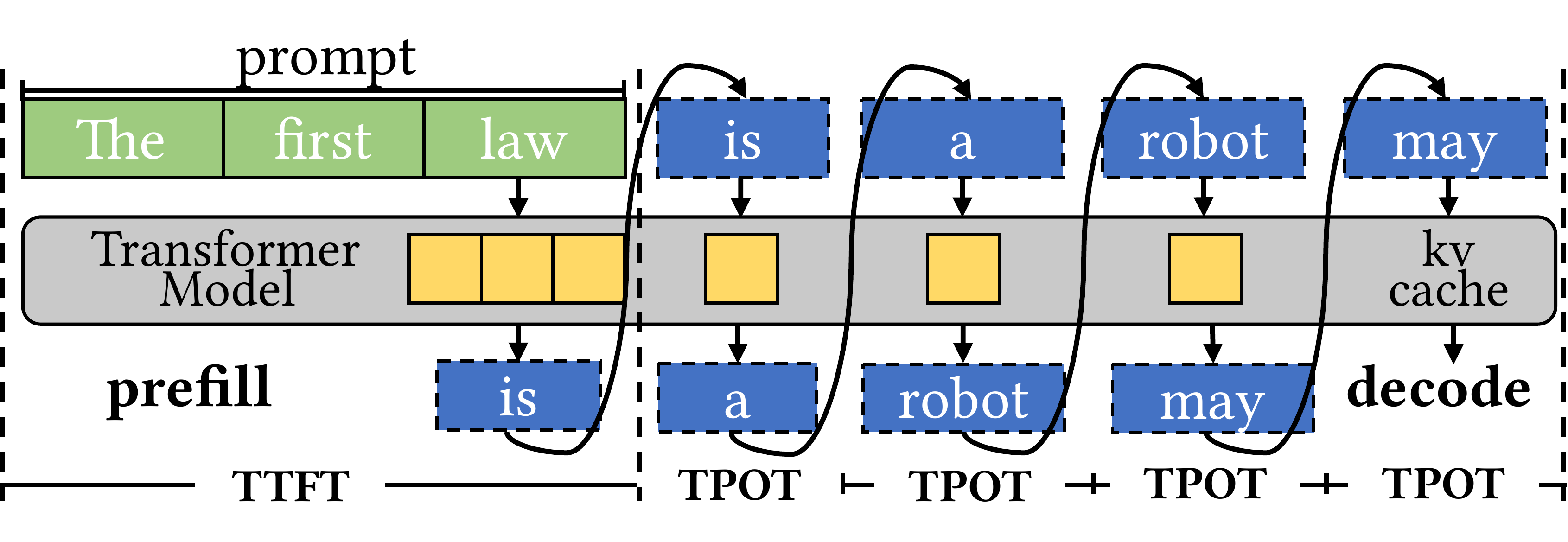}
    \caption{LLM autoregressive decoding process.}
    \label{fig:llminfer}
    \vspace{-10pt}
\end{figure}

As illustrated in Figure~\ref{fig:llminfer}, the LLM predicts the next token with the accumulated context iteratively until it encounters the end-of-sequence (EoS).
By saving the key and value embedding in the memory (i.e. KV cache), redundant computations are avoided in subsequent steps, thus the inference process is divided into prefill and decode phases.
Here we conclude their computation and memory features.

\begin{equation}
    \textbf{Q} = W_q\textbf{X}, \quad \textbf{K} = W_k\textbf{X}, \quad \textbf{V} = W_v\textbf{X}.
    \label{eq:1}
\end{equation}
\vspace{-8pt}
\begin{equation}
    Attention(\textbf{Q},\textbf{K},\textbf{V}) = softmax(\frac{\textbf{$QK^T$}}{\sqrt{d_k}})\textbf{V}
    \label{eq:2}
\end{equation}
\vspace{-5pt}
\begin{equation}
    FFN(x) = Act(x\textbf{$W_1$}+ b1)\textbf{$W_2$} + b_2
    \label{eq:3}
\end{equation}

Modern LLMs primarily adopt the Transformer architecture~\cite{vaswani2017attention}, which leverages the self-attention mechanism to model complex dependencies in sequences.
Its computation involves three main steps:
\begin{itemize}[leftmargin=*]
    \item \textbf{QKV projection} (Equation~\ref{eq:1}): Each input token is projected into query (\textbf{Q}), key (\textbf{K}), and value (\textbf{V}) embeddings.
    \item \textbf{QKV attention} (Equation~\ref{eq:2}): A weighed aggregation is performed using the scaled dot-product of queries and keys, determining how each token attends to others in the sequence. The resulting weights are applied to the value vectors, capturing contextual information.
    \item \textbf{Output projection} (Equation~\ref{eq:3}): A position-wise feed-forward layer further transforms each token's representation with a nonlinear activation, enhancing the model's capacity to learn complex patterns.
\end{itemize}

\begin{table}[!t]
\caption{Notations.}
\label{tab:hyperparameter}
\centering
\resizebox{0.9\linewidth}{!}{
\begin{tabular}{ccc}
\hline
 \textbf{Variable} & \textbf{Description} & \textbf{Notation} \\ \hline
 $prompt\_len$ & The length of prompt & S \\ \hline
 $generation\_len$ & The length of generated tokens  & G  \\ \hline
 $batch\_size$ & The number of batched requests & B \\ \hline
 $layer\_num$ & The number of model layers & L \\ \hline
 $hidden\_size$ & Input dimension of the hidden layer & H \\ \hline
 $heads$ & The number of attention heads &  M \\ \hline
 $size\_per\_head$ & The hidden state per head & D \\ \hline
\end{tabular}}
\end{table}

\begin{table}[!t]
\caption{Approximate arithmetic intensity (AI) of primary operations in LLMs. As the hidden size is usually large, negligible factors are omitted.}
\label{tab:pd_ai}
\centering
\resizebox{0.98\linewidth}{!}{
\begin{tabular}{cccccc}
\hline
\textbf{Operation }                              & \textbf{P/D}     & \textbf{FLOPS} & \textbf{Memory Access} & \textbf{Approximate AI} \\ \hline 
\multirow{2}{*}{QKV Projection}         &  Prefill & $6BSH^2$  &  $6BSH + 3H^2$   & $BS$    \\ 
                                        &  Decode  & $6BH^2$  & $6BH + 3H^2$   &  $B$             \\ \hline 
\multirow{2}{*}{Attention $QK^T$}       &  Prefill & $2BS^2H$ &  $2BSH + BS^2M$  &   $S$    \\  
                                        & Decode  & $2BSH$  & $2BSM + BH(S+1)$   & $1$          \\ \hline
\multirow{2}{*}{Attention $(QK^T)V$}    & Prefill & $2BS^2H$ & $2BSH + BS^2M$  &   $S$        \\
                                        & Decode  & $2BSH$ & $2BSM + BH(S+1)$  & $1$               \\ \hline
\multirow{2}{*}{Output Projection}      & Prefill & $2BSH^2$ & $2BSH + H^2$  &   $BS$        \\ 
                                        & Decode  & $2BH^2$ & $2BH + H^2$  & $B$          \\ \hline
\multirow{2}{*}{Dim Expansion}          & Prefill & $8BSH^2$ & $2BSH + 4H^2$  &  $BS$      \\ 
                                        & Decode  & $8BH^2$ & $2BH + 4H^2$  & $B$           \\ \hline
\multirow{2}{*}{Dim Reduction}          & Prefill & $8BSH^2$ & $2BSH + 4H^2$  &  $BS$      \\
                                        & Decode  & $8BH^2$ & $2BH + 4H^2$ & $B$       \\ \hline 
\end{tabular}}
\end{table}

\textbf{Distinct arithmetic intensity.} In Transformer-based LLM inference, matrix multiplications dominate the overall computation time, while softmax and layer normalization account for only a small fraction of the total execution time.
As listed in Table~\ref{tab:pd_ai}, there are 6 major matrix multiplication operations and we compute their arithmetic intensities separately for both prefill and decode phases, using the hyperparameters defined in Table~\ref{tab:hyperparameter}.
The arithmetic intensity is computed by dividing the total number of floating-point operations by the total amount of memory access.
Since some terms, such as $1/H$, are negligible, we omit them and present the approximate arithmetic intensity.
Although certain optimization techniques, such as FlashAttention~\cite{dao2022flashattention}, can affect the arithmetic intensity, our calculations closely reflect real-world scenarios.
As shown in Table~\ref{tab:pd_ai}, the arithmetic intensity of the prefill phase depends on both the sequence length $S$ and batch size $B$, while the decode phase primarily depends on only the batch size $B$.
Since the sequence length $S$ typically ranges from a few dozen to a few hundred tokens (or even more), the prefill phase exhibits significantly higher intensity compared to the decode phase.
Additionally, the decode phase requires loading the KV cache, further increasing memory access.
Consequently, when the two phases are executed independently, the prefill phase is generally compute-bound, whereas the decode phase is typically memory-bound.

\textbf{Memory-compute trade-off.}
During LLM inference, limited memory capacity can significantly restrict computational parallelism, making parallel inference a potential avenue for achieving superlinear speedup. A key contributor to memory consumption, beyond model weights, is the KV cache. The decode phase is typically bottlenecked by memory bandwidth, necessitating the simultaneous processing of hundreds of requests, which results in substantial memory usage. For example, in Llama-30B, the KV cache for a single token requires 1.52 MB, meaning that 128 requests with an average output length of 300 tokens would demand approximately 58.4 GB of memory, comparable to the memory footprint of model weights.
Furthermore, LLM generative tasks inherently involve variable-length sequences, where both input and output lengths are stochastic, and the output length remains unknown until inference completes. This uncertainty necessitates reserving a substantial amount of memory to prevent out-of-memory (OOM) issues, further complicating efficient resource allocation.

\subsection{LLM Batching Techniques}
To fully utilize modern GPUs, the batching technique is commonly adopted for processing deep learning workloads, where multiple samples are processed simultaneously to expose high parallelism and provide considerable hardware performance improvements.
First, given the variation in input and output lengths, continuous batching~\cite{yu2022orca} has emerged as the standard technique, enabling requests to dynamically enter or exit a batch at each iteration.
Subsequently, modern LLM serving systems typically employ either separate batching~\cite{vllm2024github} or hybrid batching~\cite{agrawal2024taming}.

Separate batching packs requests exclusively during the prefill phase and solely during the decode phase.
Given that these two phases exhibit distinct performance characteristics, i.e. compute-bound and memory-bound, the prefill phase saturates GPU computation even at a batch size of just one, while the decode phase requires a batch size in hundreds.
By contrast, hybrid batching combines requests of both prefill and decode phases, and organizes them in a hybrid batch.

\subsection{Parallel LLM Inference}

\begin{figure}[!t]
    \centering
    \includegraphics[width=0.98\linewidth]{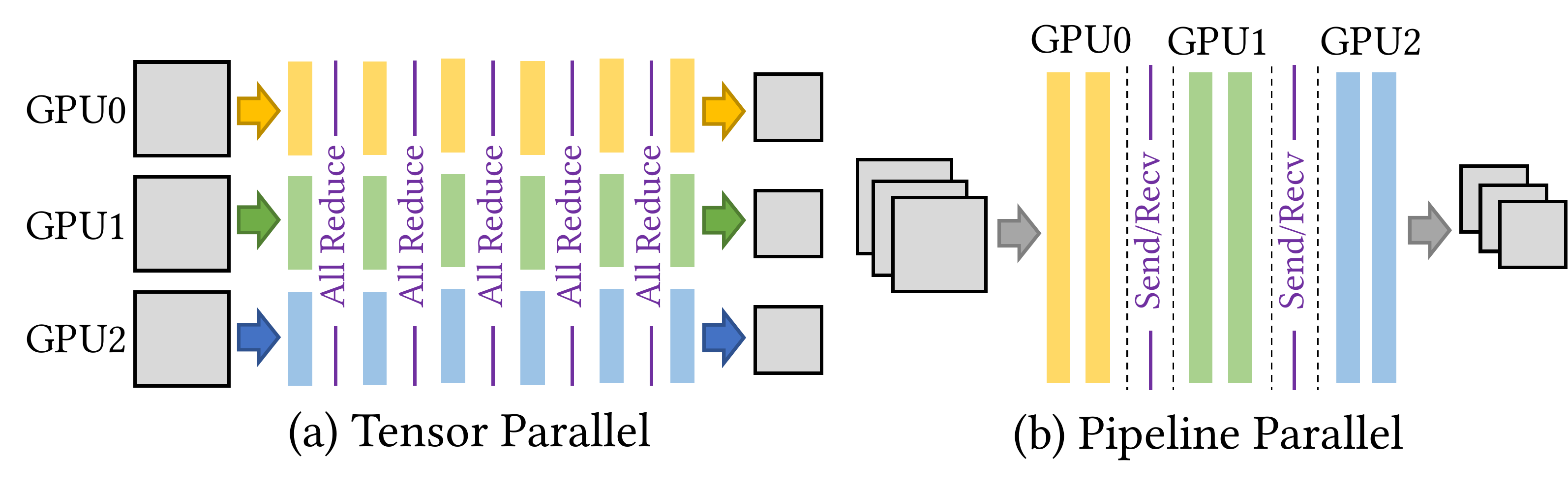}
    \caption{Tensor parallelism and pipeline parallelism.}
    \label{fig:parallel_approach}
\end{figure}


To enhance both computational and memory capacity within a single inference instance, various parallelism strategies are employed, including tensor parallelism~\cite{megatron, cheng2023atp, shazeer2018mesh} (TP), expert parallelism~\cite{singh2023hybrid} (EP), sequence parallelism~\cite{jacobs2023deepspeedulys, liu2023ring} (SP), and pipeline parallelism~\cite{gpipe2019, li2021terapipe, narayanan2019pipedream} (PP).
Since TP, EP, and SP all involve distributing the computation of a single layer across multiple devices, they exhibit a similar communication pattern. In this work, we primarily use TP as a representative example to illustrate the key concepts.

Figure~\ref{fig:parallel_approach} illustrates how a model is partitioned in the TP and PP approaches.
TP partitions each layer across multiple GPUs, with both model weights and KV cache equally distributed across GPU workers.
It demands frequent inter-device communications, with each Transformer layer involves two rounds of all-reduce operations.
In our case study of distributed TP inference using Llama-30B on four NVIDIA L20 GPUs (only PCIe), communication overhead accounts for nearly half of the total execution time.
While TP can accelerate individual inference runs, it also results in substantial idle time for computational resources due to synchronization and communication bottlenecks.

\begin{figure}[!t]
    \centering
    \includegraphics[width=0.98\linewidth]{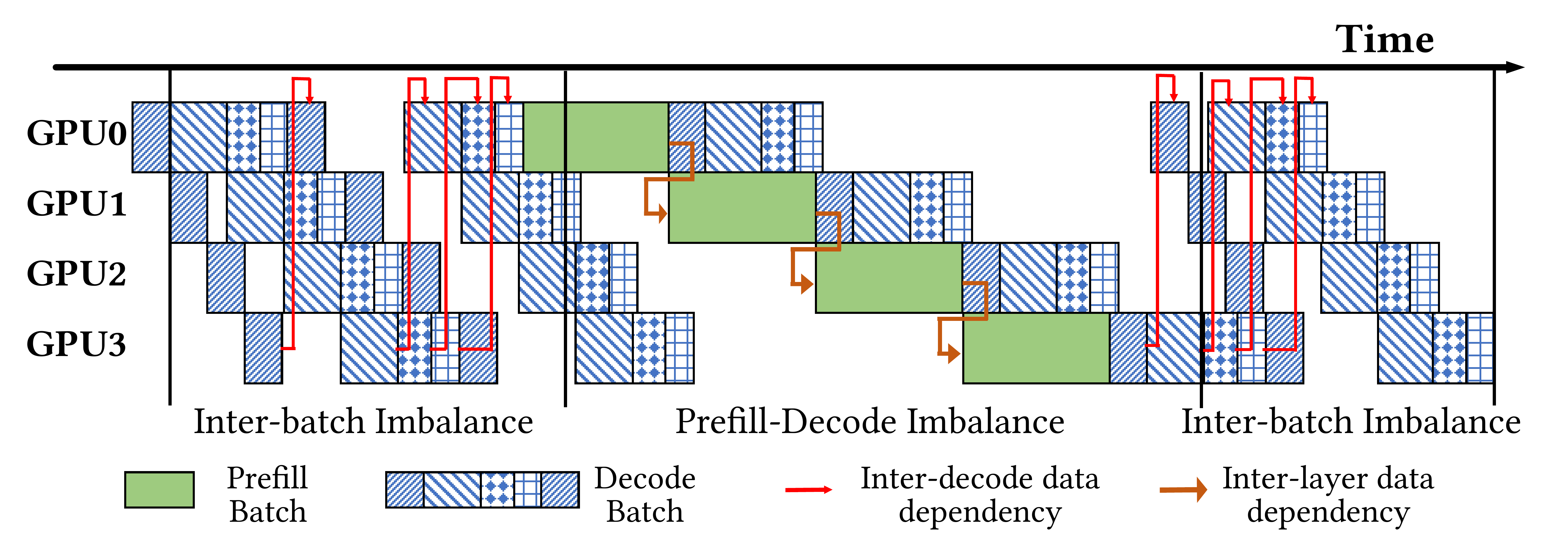}
    \vspace{-10pt}
    \caption{Pipeline bubbles.}
    \label{fig:pp_bottleneck}
    \vspace{-10pt}
\end{figure}

PP partitions a model layer-wise, where each device is responsible for a subset of layers. 
It only requires a single point-to-point communication with a much smaller data volume every few layers.
The largely-reduced communication makes it a promising approach to support LLM serving, especially on commodity hardware that high-performance interconnects like NVLINK is unavailable.
However, the imbalanced pipeline workloads and complex data dependencies of LLM inference often prevent it from being the primary choice. Figure~\ref{fig:pp_bottleneck} illustrates the execution process of the PP approach with the separate batching, which is commonly adopted in practical LLM serving systems~\cite{vllm2024github, sglang2024github}.

The imbalanced workloads come from two aspects.
First, different batch sizes of decode batches lead to different workloads, resulting in the inter-batch imbalance.
Second, the prefill-decode imbalance exists as the execution of prefill batch usually takes much longer time than the decode batch. 
Likewise, the data dependencies also come from two aspects.
Inter-layer data dependency requires data to be processed in the current layer before entering to the next layer for processing.
Inter-decode data dependency exists as the execution of generating the next token cannot start until the previous iteration completes.
Therefore, bubble problems always hinder the practical use of pipeline parallelism in serving LLMs.

\subsection{Large-scale LLM Serving}

As demand for LLM inference continues to grow, large-scale deployment and the adoption of cluster-level infrastructure have become essential to meet increasing workload requirements.
Based on whether prefill and decode phases are disaggregated, there are NoDG strategy and FuDG strategy.

\subsubsection{Non-Disaggregated Strategy}

The NoDG strategy~\cite{vllm2024github, sglang2024github, agrawal2024taming} colocates prefills and decodes in a single instance, which is responsible for the entire life-cycle of a request.
As shown in Figure~\ref{fig:twodg}(a), when a single instance is unable to handle the incoming requests, the system replicates additional instances, each functioning independently to scale out the service.
The NoDG strategy supports both separate batching and hybrid batching approaches.
As prefills often take much longer than a decoding step, when scheduling together, decodes are always delayed by the prefills, significantly elongating their TPOT; similarly, the inclusion of decodes contributes to a non-trivial increase in TTFT.
Thus, NoDG systems often suffer from low throughput as the decode phase can hardly accumulate a sufficiently large batch size to saturate the GPU under SLO constraints.
To mitigate this interference, chunked prefill~\cite{agrawal2024taming} divides a prefill request into smaller chunks and processes a prompt’s prefill phase over multiple iterations.
However, chunked prefill incurs the overhead of repeated KV cache access, and its effectiveness heavily depends on the input-to-output length ratio.



\subsubsection{Fully-Disaggregated Strategy}

As illustrated in Figure~\ref{fig:twodg}(b), the FuDG strategy~\cite{distserve, qin2024mooncake, patel2024splitwise} disaggregates the prefill and decode phases across separate instances, with the KV cache transferred between them.
When a new request arrives, it first enters a prefill instance, where the KV cache and the first token are generated. The KV cache is then transmitted to a decode instance for the remaining decoding steps.
Although it can completely eliminate the prefill-decode interference, it introduces issues related to data transmission and load imbalance~\cite{liang2025injecting}.

\begin{table}[!t]
\caption{KV cache generation speed in the prefill instance and theoretical bandwidth required for the FuDG strategy. Here each node includes 8 GPUs and tensor parallelism is applied when a single GPU's memory capacity is insufficient.}
\label{tab:kvcache}
\centering
\resizebox{0.9\linewidth}{!}{
\begin{tabular}{cccc}
\hline
 \textbf{Model} & \textbf{Device} & \textbf{Tokens/s} & \textbf{Theoretical Bandwidt}h  \\ \hline\hline
 Llama-30B & L20 & 6584.6 &  9.796 GB/s \\ \hline
 Llama-30B & A800 & 26189.2 & 38.96 GB/s \\ \hline
 CodeLlama-34B & L20 & 6838.92 & 1.25 GB/s\\ \hline
 CodeLlama-34B & A800 & 25978.88& 4.76 GB/s\\ \hline
\end{tabular}}
\end{table}

Since the FuDG strategy requires transferring massive amounts of KV cache data between prefill and decode instances, it relies on high-performance interconnects to avoid bottlenecks.
Table~\ref{tab:kvcache} presents the KV cache generation speed in a GPU node (all prefill instances), along with the the theoretical bandwidth required to transfer these KV cache data.
When deploying Llama-30B prefill instances on a node equipped with 8 NVIDIA A800 GPUs, the theoretical bandwidth required to transfer the generated KV cache data off the node exceeds 38 GB/s, necessitating at least a 400 Gbps network to sustain the throughput.
Although Grouped Query Attention (GQA)~\cite{ainslie2023gqa} in CodeLlama-34B significantly compresses KV cache size, the strategy still demands over 4.76GB/s bandwidth, requring a 50Gbps network.

Consequently, to support deployment across clusters with varying interconnect capabilities, the FuDG strategy can be further classified into intra-node FuDG (DistServe~\cite{distserve}) and inter-node FuDG (MoonCake~\cite{qin2024mooncake}), depending on whether the prefill and decode instances are colocated within the same node or distributed across different nodes.
In scenarios where inter-node interconnects are insufficient, DistServe deploys prefill and decode instances within a node and mitigates the issue by transferring KV cache data over intra-node high-speed links, such as NVLink, although these interconnects are also costly.
In contrast, MoonCake designs a centralized KV cache pool and relies on InfiniBand to connect prefill and decode instances across nodes.
In short, the FuDG strategy incurs substantial high-performance networking requirements in both intra-node and inter-node setups.

Next, the FuDG strategy also suffers from severe load imbalance issues.
Firstly, it requires careful load balancing between prefill instances and decode instances.
Due to the asymmetry in durations, adding a single prefill instance typically necessitates provisioning multiple decode instances to maintain load balance.
If prefill and decode instances are colocated in a single node due to bandwidth limitations, achieving load balance becomes somewhat infeasible.
Secondly, memory utilization across prefill and decode instances is imbalanced.
Decode instances store large amounts of KV cache, while prefill instances store much less.
As memory capacity is a valuable resource in modern LLM workloads, such imbalance may leave a significant portion of memory idle in prefill instances, resulting in suboptimal resource efficiency.

%% file: tex/Design.tex
\begin{figure*}[!t]
    \centering    \includegraphics[width=0.9\linewidth]{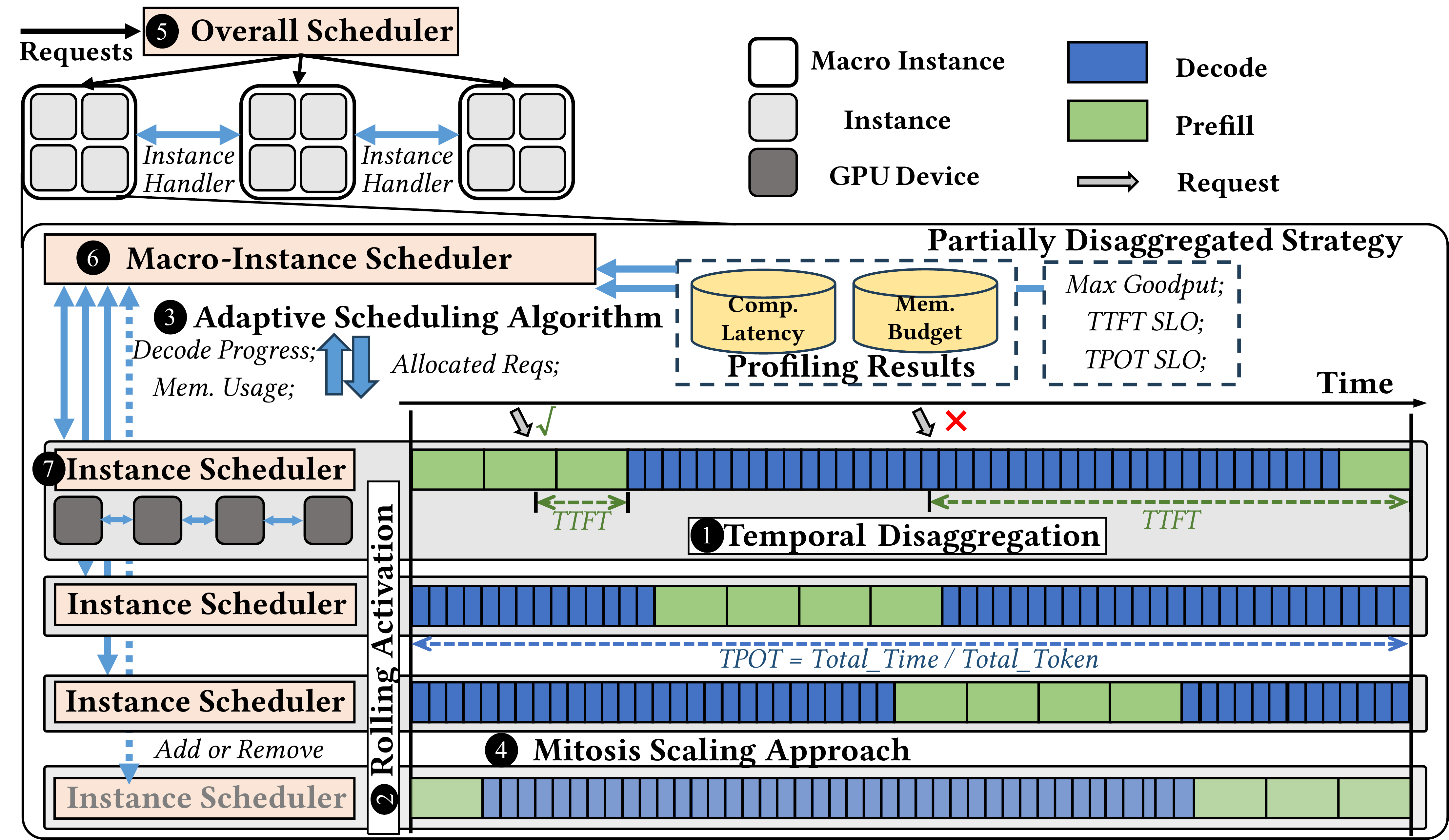}
    \caption{EcoServe Architecture Overview.}
    \label{fig:overview}
\end{figure*}

\section{EcoServe Design}

\subsection{Overview}
As depicted in Figure~\ref{fig:overview}, EcoServe employs a partially disaggregated (PaDG) strategy that proactively orchestrates both intra- and inter-instance execution to optimize goodput.
Specifically, the prefill and decode phases are disaggregated along the time dimension in each instance (\circled{1} Temporal Disaggregation), while coordination is conducted across different instances to ensure continuous availability (\circled{2} Rolling Activation).

The EcoServe system is organized in a hierarchical architecture, comprising three levels of scheduling: the overall scheduler, the macro-instance scheduler, and the instance scheduler.
The instance scheduler (\circled{5}) is responsible for managing execution within a single instance, including coordinating prefill and decode phases, orchestrating multiple devices, and executing the directives from higher-level schedulers. 
The macro-instance scheduler coordinates multiple instances by aggregating their execution states and dispatching requests to appropriate instance according to given profiling results and service-level objectives (SLOs), where the macro instance (\circled{6}) is defined as a unique abstraction level introduced in EcoServe and serves as the smallest scheduling unit in the system.
The overall scheduler (\circled{7}) dispatches requests to macro-instances based on their capabilities.
Besides, it manages the capacity scaling in different macro instances, such as transferring instance handler between macro instances.
In this work, we primarily focus on the internal architecture within a macro-instance.

Furthermore, EcoServe integrates the adaptive scheduling algorithm and the mitosis scaling approach to enable practical implementation of the PaDG strategy.
The adaptive scheduling algorithm (\circled{3}) is applied at both the instance scheduler and macro-instance scheduler and makes them coordinated.
From the perspective of the instance scheduler, it executes decodes while accumulating sufficient slack to safely admit new requests, until it receives a continuous stream of incoming requests from the macro-instance scheduler.
From the perspective of the macro-instance scheduler, it continuously receives new requests from the overall scheduler and updates execution status from individual instances.
Based on this information, it estimated which instance and how many prefill tokens should be forwarded under given constraints.


The mitosis scaling approach (\circled{4}) enables fine-grained capacity adjustments to accommodate fluctuations in LLM inference workloads over extended periods of time.
Since the smallest scheduling unit in EcoServe is the macro instance, adjusting capacity at the granularity of an entire macro instance can be inflexible and often leads to resource under utilization or waste.
Inspired by biological cell mitosis, this scaling strategy incrementally adds or removes instances within a macro instance in response to changed demand. 
Once predefined thresholds are reached, it then splits a macro instance or merges two macro instances, enabling elastic adaptation to workload changes with fine-grained control.
Further, to avoid instance re-initialization and execution interruption, a serializable proxy object is designed for flexible instance migration between macro instances.

\subsection{Partially Disaggregated Strategy}


Figure~\ref{fig:overview} contains the proactive intra-instance and inter-instance scheduling of the PaDG strategy.
The proactive intra-instance scheduling, referred to as temporal disaggregation, reduces prefill-decode interference and enhance throughput, though at the cost of increased TTFT.
The proactive inter-instance scheduling, guided by rolling activation, plays a key role in rescuing TTFT, ensuring timely processing of new requests.

\subsubsection{Temporal Disaggregation}
\label{Temporal-Disaggregation}
To mitigate prefill-decode interference, the PaDG strategy disaggregates the prefill and decode phases along the temporal dimension within each instance.
Unlike the FuDG strategy, which assigns distinct prefill and decode roles to different instances, PaDG assigns these roles to different time slots within the same instance.
This design preserves execution locality while mitigating prefill-decode interference, thereby achieving efficient on-device resource utilization and eliminating cross-instance data transmission.

Through proactive intra-instance scheduling, each inference instance processes only one phase type at a time for an extended duration.
As illustrated in Figure~\ref{fig:overview}, when a new request arrives at an instance that is currently processing prefills, it can be immediately processed, thereby meeting the TTFT SLO.
However, if the target instance is currently performing the decode phase, the request must wait until the instance transitions to the prefill phase, leading to an unacceptable increase in TTFT.
Consequently, since each phase occupies an instance for an extended period, this intra-instance scheduling significantly degrades TTFT, making it difficult to meet the corresponding TTFT SLO.
In comparison, modern LLM serving systems typically render outputs in a typewriter mode, where the TPOT SLO can be satisfied as long as a sufficient number of tokens are generated within a time window.
This means that if the decode execution is faster than the TPOT constraint, it can accumulate spare time (referred to as saved TPOT), which can be used for interruptions in the decode phase without violating the SLO.


\subsubsection{Rolling Activation}
Although a single instance can only remain in either the prefill or decode phase at any given time, and therefore cannot immediately process newly arrived requests, rolling activation proactively schedules multiple instances and staggers their prefill phases over time, thereby ensuring the continuous availability of prefill processing.
As illustrated in Figure~\ref{fig:overview}, these instances are activated to perform prefill phases in a cyclic pattern.
From the perspective of individual requests, they are always routed to instances that are currently in the prefill phase and can be processed intermediately.
In this way, rolling activation reduces waiting time and rescues TTFT.
Since the output length is undetermined, instances require to constantly update their statuses, such as decode progress and memory usage, to macro instance for coordination. 

\subsection{System Metrics v.s. User Experience}
\label{metrics}
\begin{figure}[!t]
    \centering
    \includegraphics[width=0.95\linewidth]{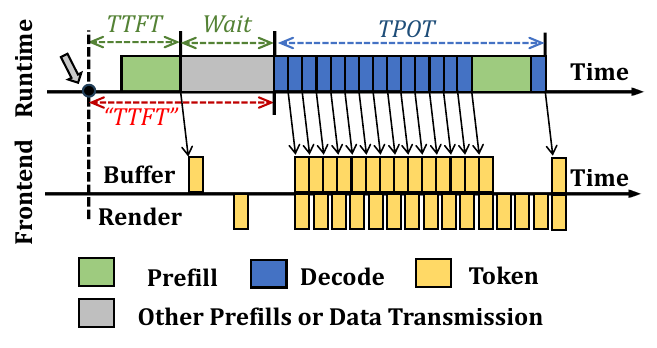}
    \caption{Runtime and frontend Timing.}
    \label{fig:metrics}
\end{figure}

Figure~\ref{fig:metrics} demonstrates the timing characteristics at both the runtime and frontend.
Once a token is generated, it is transmitted to the frontend, where it is buffered before being rendered.
From the runtime perspective, classical metrics such as Time to First Token (TTFT) and Time Per Output Token (TPOT) are commonly used to characterize the latency behavior of a request's prefill and decode phases, respectively.
However, for existing LLM serving systems, they have been insufficient to reflect the service quality.

TTFT and TPOT alone are insufficient to capture the performance characteristics of the prefill-decode switching phase.
Before a request enters its decode phase, additional operations occur across all strategies, NoDG, PaDG, and FuDG.
For the NoDG and PaDG strategies, prefills from other requests may occur before a given request can enter its decode phase.
Similarly, the FuDG strategy incurs KV cache transmission overhead prior to the decode phase.
Thus, the phase-switching waiting time should be introduced to provide a more comprehensive evaluation of LLM serving systems.
This metric has implicitly appeared in previous work, and it is frequently misrepresented, potentially masking key limitations in existing LLM serving systems.

To maintain consistency with previous studies, we continue to use TTFT as the metric for evaluating prefill latency.
However, in this context, the reported TTFT actually encompasses two components: the true TTFT and the phase-switching waiting time.
Notably, this definition of TTFT represents \textbf{a stricter SLO}, as it includes additional overhead within the same latency constraint.
Accordingly, the measurement of TPOT begins after the phase-switching delay.

\subsection{Adaptive Scheduling Algorithm}
\label{sec:algo}

\begin{algorithm}[!t]
\setlength{\textfloatsep}{0.5cm}
\setlength{\floatsep}{0.5cm}
\setlength{\intextsep}{-5em}
\KwData{current request: $\mathit{req}$; instance list: $\mathit{instances}$;}
\SetKwFunction{InterSchedule}{InterSchedule}
\SetKwFunction{CheckConstraints}{CheckConstraints}
\SetKwProg{Fn}{Function}{:}{}
\Fn{\InterSchedule{$\mathit{req}$}}{
     $\text{prev\_idx} \leftarrow$ last request's routed instance\;
     $\text{last\_instance} \leftarrow \mathit{instances[prev\_idx]}$\;
    \If{\CheckConstraints($\mathit{last\_instance,} \text{req}$)}{
        route $\mathit{req}$ to $\mathit{instance[prev\_idx]}$\;
    }
    \Else{
     $\text{next\_idx} \leftarrow \mathit{(prev\_idx + 1) \% len(instances)}$  \;
        route $\mathit{req}$ to $\mathit{instance[next\_idx]}$\;
    }
}
\caption{Inter-Instance Scheduling Algorithm}
\label{algo:inter}
\end{algorithm}

\begin{algorithm}[!t]
\setlength{\textfloatsep}{0.5cm}
\setlength{\floatsep}{0.5cm}
\setlength{\intextsep}{-5em}
\KwData{System constraints: $\mathit{SLO_{TTFT}}, \mathit{SLO_{TPOT}}$;}
\SetKwFunction{CheckConstraints}{CheckConstraints}
\SetKwProg{Fn}{Function}{:}{}
\Fn{\CheckConstraints{$\mathit{instance}$, $\mathit{req}$}}{
    \textbf{Constraint 1: TTFT} \\
    $t_{\text{switch}} \leftarrow$ phase switching timestamp\; 
    $\mathit{pending\_prefills} \gets \bigl\{ r \in instance.reqs\mid 
     r.arrival\_time \geq t_{\text{switch}} \bigr\} \cup \{req\}$

     $prefill\_times \leftarrow $ predict $\mathit{pending\_prefills}$ durations \;
    $t_\text{total} \leftarrow \sum{prefill\_times}$\;
    \If{$t_\text{total}$ > $\mathbf{SLO_{TTFT}}$}{
        return NotSatisfied\;
    }
    \textbf{Constraint 2: TPOT} \\
    $\mathit{existed\_decodes} \gets \bigl\{ r \in instance.reqs\mid 
     r.arrival\_time < t_{\text{switch}} \bigr\}$\;
     $saved\_tpots \leftarrow []$\;
     $current\_time \leftarrow $ current timestamp\;
    \ForEach{$r \in existed\_decodes$}{
        $L \leftarrow r.output\_length$\;
        $saved\_tpot \leftarrow L \times SLO_{TPOT} - (current\_time - r.first\_token\_time)$
        $saved\_tpots$.append($saved\_tpot$) 
    }
    $mean\_saved\_tpot \leftarrow mean(saved\_tpots)$\;
    \If{$mean\_saved\_tpot < t_\text{total}$}{
        return NotSatisfied\;
    }
    \textbf{Constraint 3: KV Cache capacity} \\
    \If{$req\_kvcache\_size > remain\_memsize$}{
        return NotSatisfied\;
    }
    return Satisfied
}
\caption{Constraint Checking Algorithm}
\label{algo:check}
\end{algorithm}

        



To enable proactive scheduling within and across instances, we propose an adaptive scheduling algorithm consisting of three sub-algorithms: the Inter-Instance Scheduling Algorithm, the Intra-Instance Scheduling Algorithm, and the Constraint Checking Algorithm, coordinating decisions at multiple levels. 

The inter-instance and the intra-instance scheduling algorithms guide the fundamental execution of the instance macro-scheduler and the instance scheduler.
In general, the macro-instance scheduler and the instance scheduler function in a master-slave manner.
From the instance scheduler’s perspective, although the final outcome is that prefill and decode phases are disaggregated along the temporal dimension, its scheduling algorithm prioritizes prefills.
It continues processing active decodes, periodically updating its progress to the macro-instance scheduler, and switches to prefills upon receiving new requests from the macro-instance scheduler.

From the macro-instance scheduler's perspective, it receives status updates from all instances and schedules them to achieve rolling activation.
Specifically, the inter-instance scheduling algorithm traverses all instances and routes requests cyclically.
As shown in Algorithm~\ref{algo:inter}, for an incoming request, the algorithm first attempts to route it to the same instance that processed the previous request.
If the selected instance meets the constraints verified by the Constraint Checking Algorithm, the request is forwarded to this instance. 
If the instance cannot satisfy the constraints, the algorithm will check the next available instance.

The constraint checking algorithm, as described in Algorithm~\ref{algo:check}, is responsible for verifying that assigning an incoming request to an instance will not violate the TTFT/TPOT latency SLOs or exceed the instance's available memory capacity.
First, the algorithm ensures that the total duration  (denoted as $t_{\text{total}}$) will not exceed the $SLO_{TTFT}$ after adding a new request to the pending prefills during this prefill phase.
The prefill duration of a single request can be predicted in advance by profiling sequences of various lengths.
This ensures the TTFT constraint, as outlined in Section \ref{metrics}. 
Additionally, by utilizing updated information from the instance, the algorithm calculates the saved TPOT by subtracting the required time from the achieved time.
As discussed in~\ref{Temporal-Disaggregation}, provided that $t_{\text{total}}$ does not exceed the saved TPOT, the TPOT constraint will be satisfied.
Finally, the algorithm ensures that the combined KV cache size of the requests does not exceed the remaining GPU memory capacity, preventing memory overflow during processing.

\subsection{Mitosis Scaling Approach}

\begin{figure}[!t]
    \centering
    \includegraphics[width=0.95\linewidth]{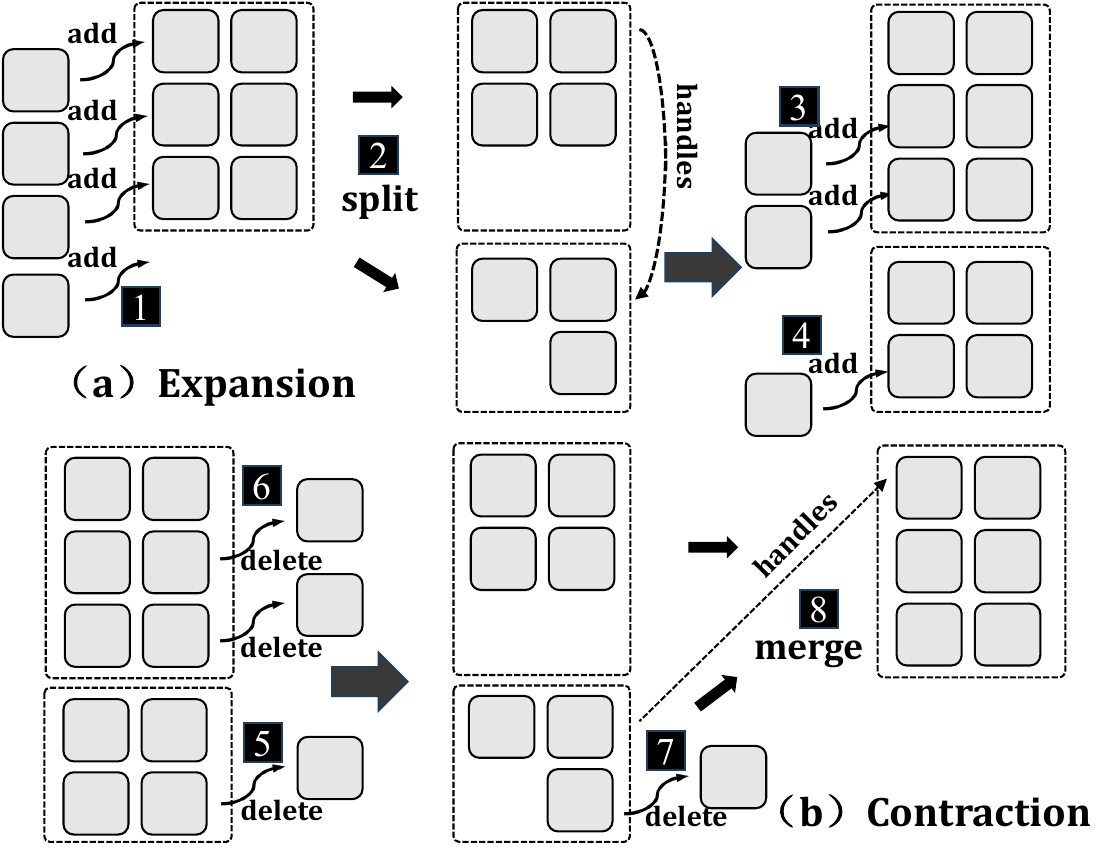}
    \caption{The illustration of the expansion and contraction processes. Here $N_l = 3$ and $N_u=6$.}
    \label{fig:scaling}
\end{figure}

Although the macro instance is the smallest scheduling unit in EcoServe, scaling capacity at this granularity is often inefficient and inflexible.
To address this limitation, the mitosis scaling approach, inspired by biological cell mitosis, provides a solution to adjust capacity at the instance level, allowing EcoServe to adapt more precisely to workload fluctuations.
In general, this strategy first adds or removes instances within a macro instance, and subsequently adds or removes entire macro instances through splitting or merging.



\subsubsection{Expansion and Contraction.}
We initially set two hyperparameters, $N_l$ and $N_u$, representing the lower and upper bounds on the number of instances in a macro instance.
If all instances are symmetry in network topology, the number of instances within a macro instance can theoretically range from 1 to infinity.
However, a small $N_l$ may lead to frequent phase switching, while a large $N_u$ can introduce instance management overhead and potentially become a scheduling bottleneck.

Figure~\ref{fig:scaling} illustrates an example of how the expansion and contraction processes are performed within the system.
Here the scaling can be triggered either when the system fails to meet the defined SLOs or when there is sustained resource underutilization.
New instances are incrementally added until the number of instances exceeds the upper limit $N_u$, at which point a new macro instance containing $N_l$ instances is split off from the original macro instance (step \squared{2}).
If additional instances are still required, they are first added to the original macro instance until it again reaches $N_u$ (step \squared{3}), and subsequent instances are then added to the new macro instance (step \squared{4}).

On the contrary, when the capacity becomes excessive and the contraction process is triggered, instances are firstly removed from the smallest macro instance until the number of instances in the macro instance reaches $N_l$ (step \squared{5}).
Next, instances start to be removed from a full macro instance (step \squared{6}).
When the total number of instances across these two macro instances reaches $N_u$ (step \squared{7}), they will be merged into a single macro instance after one additional instance is removed (step \squared{8}).
After expansion or contraction, each macro instance continues scheduling according to the adaptive algorithm, with no additional logic required.
Thus, the system generally maintains multiple full macro instances, along with one or two partially filled macro instances.

\subsubsection{Flexible Instance Migration.}
To dynamically split or merge macro instances without reinitializing or interrupting individual instances, we design a serializable proxy object that enables instance handles to be transferred between different macro-instance schedulers.
At the core of this design is the \textit{InstanceHandler} metadata, which encapsulates essential information such as the actor ID, worker address, function calls, and other relevant attributes.
When a handler is transferred between macro-instance schedulers (i.e., across processes), it is first serialized using the pickle library and then sent to the target macro-instance scheduler.
The transmission process is coordinated by the overall scheduler. 
Upon deserialization, the receiving process reconstructs a fully functional proxy, which can issue function calls through the RPC-like system.
This design enables logical migration of instances across macro-instance schedulers without interrupting their execution, thereby supporting more flexible and low-overhead scaling.

%% file: tex/Evaluation.tex
\section{Evaluation}

EcoServe utilizes vLLM as the single-device runtime, with Ray controlling multiple devices per instance through RPC-like control, while ZeroMQ facilitates synchronization across instances in the macro-instance scheduler.
We evaluate EcoServe across LLMs of varying sizes and on diverse application datasets and clusters.

\subsection{Experimental Setup}
\textbf{Cluster testbed.} We conduct our experiments on two clusters. The primary testbed is a production-level cluster deployed within a technology company, representing a typical infrastructure setting in modern data centers.
This cluster consists of 8 nodes with a total of 64 GPUs, each node equipped with 8 NVIDIA L20-48GB GPUs connected via PCIe only.
These nodes are interconnected through a standard 10Gbps Ethernet.
The second testbed consists of two nodes, each equipped with 8 NVIDIA A800-80GB GPUs, with all GPUs connected via PCIe only.
Unlike the primary cluster, this setup features a higher-bandwidth interconnect between nodes, i.e. 25Gbps RoCE.

\begin{table}[!t]
\caption{Dataset Features and Corresponding SLOs.}
\label{tab:application}
\centering
\resizebox{0.95\linewidth}{!}{
\begin{tabular}{ccccccc}
\hline
\textbf{DataSet} &  \textbf{$In_{\text{Avg}}$} & \textbf{$In_{\text{Med}}$} & \textbf{$Out_{\text{Avg}}$} & \textbf{$Out_{\text{Med}}$} & \textbf{$SLO_{\text{TTFT}}$} & \textbf{$SLO_{\text{TPOT}}$}   \\ \hline
Alpaca-gpt4   & 20.63 & 17.00 & 163.80 & 119.00 & 1s & 100ms     \\ \hline
ShareGPT  & 343.76 & 148.00 & 237.20 & 152    & 5s  & 100ms     \\ \hline
LongBench  & 2686.89 & 2736.50 & 101.78 & 19  &  15s &  100ms   \\ \hline
\end{tabular}}
\vspace{-10pt}
\end{table}

\textbf{Model, dataset and workloads setup.}
We choose three representative LLM models, i.e. Llama-30B~\cite{touvron2023llama}, CodeLlama2-34B~\cite{2024codellama}, and Qwen2-72B~\cite{bai2023qwentechnicalreport}, in our experiments.
LLaMA-30B adopts the standard multi-head attention (MHA) mechanism, while CodeLlama2-34B and Qwen2-72B employ the emerging grouped-query attention (GQA) mechanism~\cite{ainslie2023gqa}.
By sharing keys and values across multiple query heads, GQA significantly reduces KV cache size during inference, thereby alleviating the transmission overhead associated with the FuDG strategy.
We use BF16 precision in all experiments.

For target applications and corresponding datasets, as in Table~\ref{tab:application}, we select three representative applications with diverse input and output length distributions and remove outlier samples by truncating inputs to a maximum length of 4096, following prior studies~\cite{distserve, qin2024mooncake, kwon2023efficient}.
\begin{itemize}[leftmargin=*]
    \item \textbf{Alpaca-gpt4:} This dataset is used for the human instruction application. As shown in Table~\ref{tab:application}, it is characterized by short input sequences and relatively long outputs, with the average output length approximatebly 10 times longer than the input length.
    \item \textbf{ShareGPT:} The dataset refers to the chatbot application, featuring relatively balanced input and output lengths.
    \item \textbf{LongBench:} This dataset is used for the summarization application, where the goal is to generate a concise summary for a long article. As a result, it is characterized by long input sequences and short outputs. 
\end{itemize}
We set TTFT and TPOT SLOs based solely on applications, without differentiating between model sizes.
Our SLOs are, in most cases, stricter than those in prior works~\cite{distserve, qin2024mooncake}.
We pair each model with each dataset to construct multiple alternative workloads.
To emulate realistic serving, a Poisson distribution is applied to a fixed request rate to introduce minor fluctuations.

\textbf{Baseline.}
We compare EcoServe against four baseline systems in NoDG or FuDG strategies. We follow their released implementations, and luckily, all baselines are built with vLLM~\cite{vllm2024github} as the underlying runtime, ensuring fairness.
\begin{itemize}[leftmargin=*]
    \item \textbf{vLLM~\cite{vllm2024github}:} This is the NoDG strategy with separate batching and prefill-priority scheduling, which is originally provided by vLLM system. 
    \item \textbf{Sarathi~\cite{agrawal2024taming}:} This is the NoDG strategy with hybrid batching, decode-priority scheduling, and the chunked prefill technique.
    \item \textbf{DistServe~\cite{distserve}:} The intra-node FuDG strategy that prefill and decode instances colocate in the single node. Notably, while DistServe includes a strategy that distributes each instance across nodes and limits the KV cache transmission within a node, this strategy is only applied to pipeline parallelism and cannot satisfy SLOs in our setting.
    \item \textbf{MoonCake~\cite{qin2024mooncake}:} 
    The inter-node FuDG strategy that prefill and decode instances can be assigned to different nodes. MoonCake introduces an intermediate KV cache pool, which acts as a centralized buffer for KV cache transmission. Even when the prefill and decode instances reside on the same node, KV cache still needs to be transferred through this intermediate pool. To mitigate the load imbalance issue, we perform different P/D ratio and select the optimal one. 
\end{itemize}

\begin{figure*}[!t]
    \centering    \includegraphics[width=0.99\linewidth]{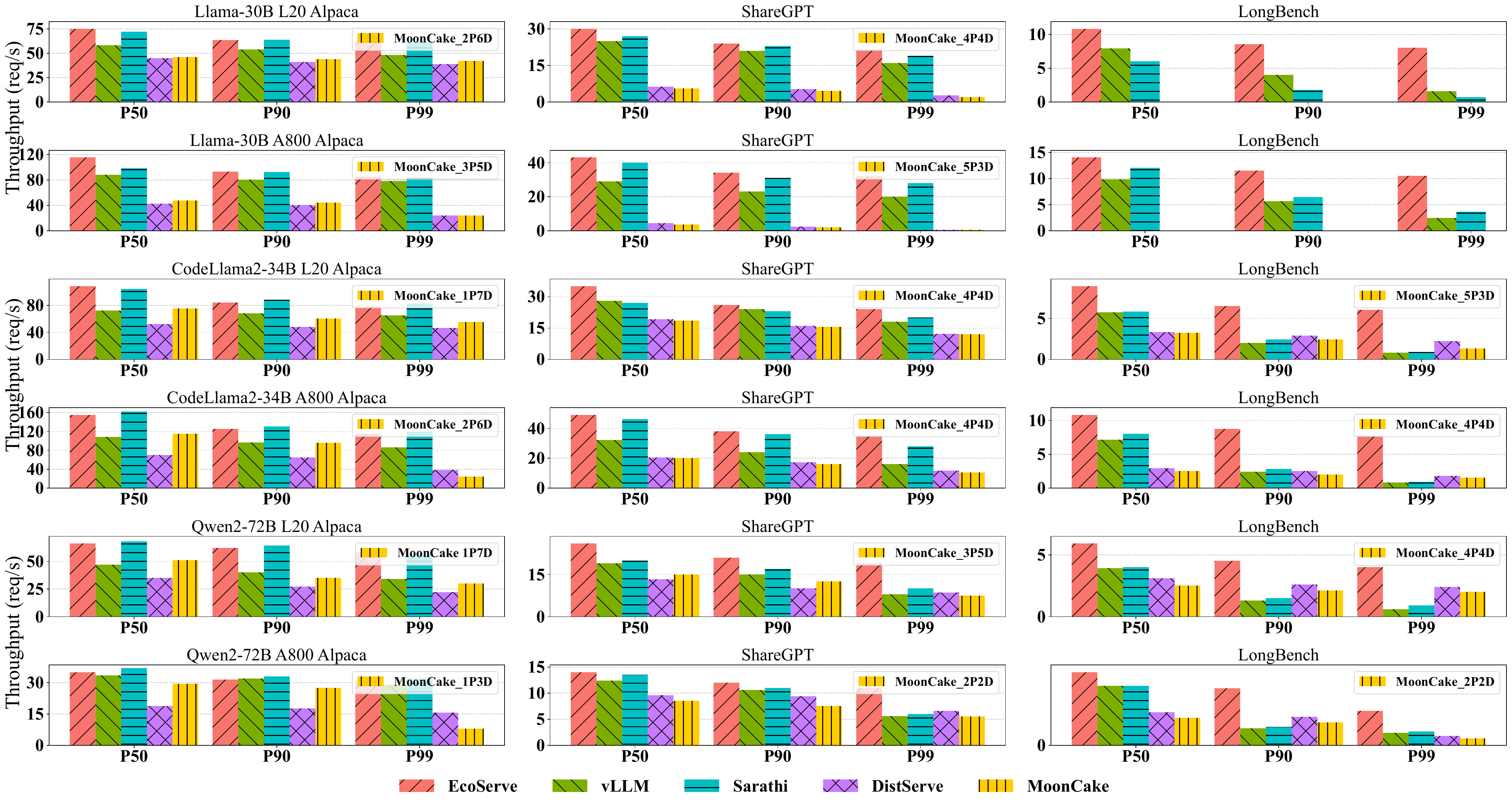}
    \caption{End-to-end performance comparison. MoonCake and DistServe cannot meet SLOs in Llama-30B with LongBench.}
    \label{fig:total_experiments}
\end{figure*}

\textbf{Metrics.}
We also use SLO attainment as the evaluation metric following prior works.
Specifically, we compare system throughput under different levels of SLO attainment, including the P50, P90, and P99 percentiles.
The throughput is collected by incrementally increasing the request rate until the system fails to reach the attainment.

\subsection{End-to-end Performance Evaluation}
We compare EcoServe against baselines across full combinations.
For the L20 cluster, we employ 32 GPUs in 8 nodes and configure the models with tensor parallelism (TP):
Llama-30B and CodeLlama2-34B are both run with TP=4, while Qwen2-72B is configured with TP=8.
To alleviate the bandwidth limitations in MoonCake, we deploy a single instance per node, reducing the bandwidth contention, and there are 8 instances for each model.
For the A800 cluster, we use 16 GPUs and configure LLaMA-30B and CodeLlama2-34B with TP=2, and Qwen2-72B with TP=4.
Accordingly, LLaMA-30B and CodeLlama2-34B each have 8 instances, while Qwen2-72B has 4 instances.

\textbf{Overall Comparison.}
In Figure~\ref{fig:total_experiments}, EcoServe outperforms  baselines in most cases.
For NoDG systems, EcoServe achieving an average P90 goodput improvement of 83.76\% over vLLM and 71.97\% over Sarathi.
By mitigating the prefill-decode interference, the PaDG strategy provides a larger room for balancing TTFT and TPOT through cross-instance cooperation.
NoDG systems can still achieve comparable or even slightly better performance than EcoServe, such as when serving the Alpaca dataset.
Since the Alpaca dataset features very short input lengths, and SLOs are already loose enough that the extra trade-off space offered by PaDG becomes less impactful.
For FuDG systems, although FuDG systems can deliver performance comparable to or better than NoDG systems when serving models with reduced KV cache and datasets with relatively long outputs, they fall significantly behind EcoServe.
EcoServe achieves an average P90 goodput improvement of 192.41\% over DistServe and 218.22\% over MoonCake.

\textbf{Comparison Across SLO Attainment Levels.}
All systems experience a decline in throughput as the SLO attainment level increases from P50 to P99.
However, EcoServe demonstrates higher tolerance in tighter SLOs.
At P50 SLO attainment, EcoServe achieves 36.49\%, 19.82\%, 180.73\%, and 194.62\% higher throughput compared to the baselines.
Under the tighter P90 SLO attainment, these improvements increase significantly to 83.76\%, 71.97\%, 192.41\%, and 218.22\%.
Next, this gap further widens, and some baseline systems are unable to meet the P99 SLO attainment.
This validates that PaDG can provide a larger room for balancing TTFT and TPOT through inter-instance cooperation.

\textbf{Comparison Across Models.}
EcoServe averagely outperforms NoDG systems' throughput by 65.00\%, 83.30\%, and 85.30\% for serving Llama-30B, CodeLlama2-34B, and Qwen2-72B models under P90 SLO attainment, demonstrating consistent performance improvements across models.
In contrast, when compared to FuDG systems, EcoServe’s advantage under P90 SLO attainment varies significantly across models, achieving 507.67\%, 125.45\%, and 83.61\% throughput improvements respectively.
Llama-30B presents significant performance degradation in FuDG systems, primarily due to its larger KV cache size, while CodeLlama2-34B and Qwen2-72B use the emerging GQA~\cite{ainslie2023gqa} with reduced size, thereby alleviating transmission overhead.
Compared to CodeLlama2-34B, Qwen2-72B performs better.
Since the computational cost grows quadratically with model size, Qwen2-72B has relatively smaller KV cache.

\textbf{Comparison Across Clusters.}
Under P90 SLO attainment, EcoServe achieves an average throughput improvement of 71.41\% over NoDG systems and 285.78\% over FuDG systems on the A800 cluster, and 84.33\% over NoDG systems and 124.86\% over FuDG systems on the L20 cluster.
The A800 and L20 clusters exhibit a similar performance trend when compared to NoDG systems, but show notable differences in their comparison with FuDG systems.
Although the A800 cluster is equipped with higher bandwidth, it appears less favorable for FuDG systems.
As shown in Table~\ref{tab:kvcache}, while the bandwidth increases by 2.5×, the processing capability improves by over 4×, thereby making the inter-node network an even more significant bottleneck.

\textbf{Comparison Across Applications.}
Under P90 SLO attainment, EcoServe achieves an average throughput improvements of 10.44\%, 20.60\%, and 202.57\% over NoDG on the Alpaca, ShareGPT, and LongBench datasets, while outperforming FuDG by 74.80\%, 363.10\%, and 164.42\% on these datasets.
The improvement over FuDG on the LongBench would be higher, as we exclude its performance on Llama-30B due to execution failures.
Comparing with NoDG systems, shorter input lengths result in reduced prefill-decode interference and fewer repeated accesses to KV cache during chunked prefill, allowing EcoServe to perform better.
In the case of FuDG, datasets with longer input and relatively shorter outputs require more prefill instances to generate KV cache, increasing network transmission pressure and leading to worse performance.

\subsection{Scaling Capability}
This section evaluates the scaling efficiency of EcoServe.
First, we double the number of instances and assess its goodput capability.
Next, we evaluate the fine-grained scaling ability under dynamically incremental request rate.

\begin{figure}[!t]
    \centering    \includegraphics[width=0.9\linewidth]{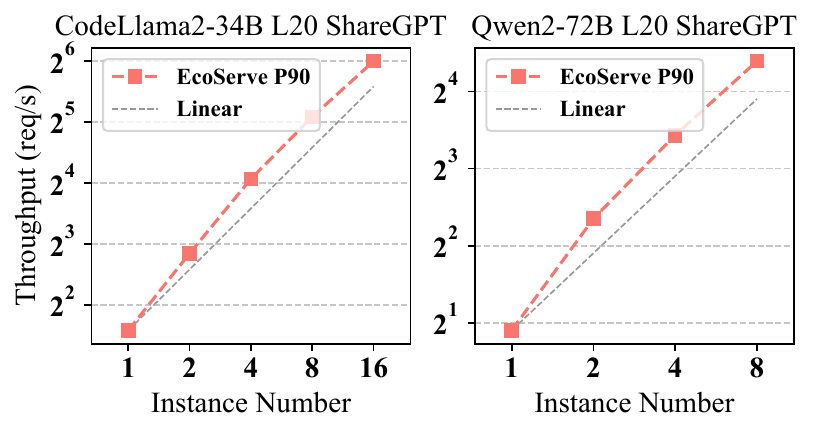}
    \caption{Static coarse-grained scaling.}
    \label{fig:coarse_scaling}
\end{figure}

\subsubsection{Static Coarse-grained Scaling}
This section evaluates the static scaling as the available resources are increased by a factor of two.
The CodeLlama2-34B and Qwen2-72B models are tested on the L20 cluster, using TP=4 for CodeLlama2-34B and TP=2 for Qwen2-72B, with the ShareGPT dataset as the workload.

Figure~\ref{fig:coarse_scaling} demonstrates that both models' serving achieve superlinear improvement under P90 SLO attainment.
For instance, when scaling from 1 instances (4 GPUs) to 4 instances (16 GPUs), the CodeLlama2-34B serving achieves 5.6$\times$ throughput under P90 SLO attainment.
First, this can be attributed to the fact that EcoServe incurs minimal overhead in managing more instances within a macro instance and the nodes in this cluster are symmetrical in topology.
More importantly, adding more instances lead to more space for mitigating inter-phase interference, enabling higher arithmetic intensity and better GPU saturation.
Assuming a macro instance contains only a single instance, the PaDG strategy actually degrades to the NoDG strategy, and two phases still switch frequently and interference severely in a single instance.
This superlinear scaling effect will plateau once a sufficient number of instances is reached.

\begin{figure*}[!t]
    \centering    \includegraphics[width=0.95\linewidth]{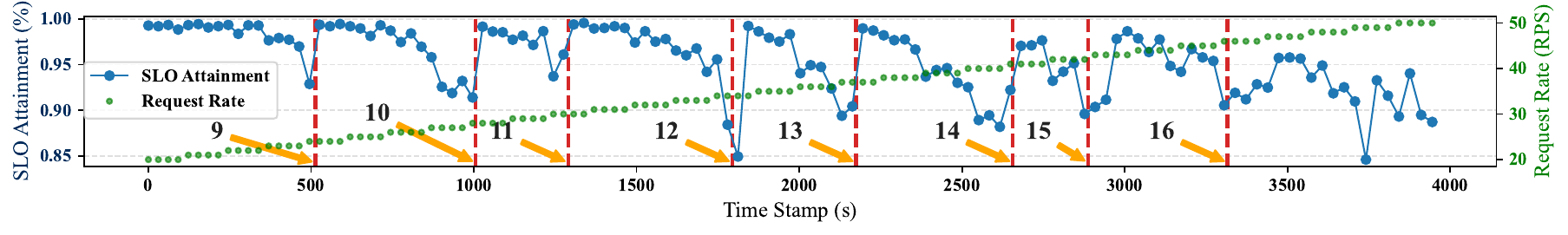}
    \caption{Dynamic fine-grained scaling. Individual instances are dynamically added to a macro instance as request rates increase. Here $N_l = 4$ and $N_u=16$.}
    \label{fig:fine_scaling}
\end{figure*}

\subsubsection{Dynamic Fine-grained Scaling}

This section evaluates the fine-grained scaling, where individual instances are incrementally added to a macro instance as request rates increase.
We use CodeLlama2-34B on the L20 cluster with TP=4, and the ShareGPT dataset as the workload.
The request rate is gradually increased every 2 minutes, ranging from 20 to 50 requests per second, and SLO attainments are collected every 30 seconds.
Based on the weak scaling experiments, we set the hyperparameters to $N_l = 4$ and $N_u=16$, as division of the macro instance would lead to performance degradation.
The system starts with 8 instances and finally uses up all GPUs.
As shown in Figure~\ref{fig:fine_scaling}, increasing the request rate initially results in a drop in SLO attainment, which is then restored by the addition of a new instance.
The adaptive scheduling algorithm can immediately route new requests to the newly added instance, leaving more time slots for existing instances to process decodes.

Although the node number does not necessarily trigger macro instance splitting and instance migration, we further assess the serializable proxy object in the mitosis scaling approach when serving at larger scale.
The use of the serializable proxy object ensures that the migration process does not interrupt instance execution, introducing less than 100 ms of overhead.
This overhead can be entirely hidden by triggering the migration during the decode phase.
In contrast, interrupting and re-initializing an instance incurs much higher overhead.
For instance, re-initializing CodeLlama2-34B from L20 node's local storage takes about 3 minutes, and this can be further prolonged when weights are loaded from remote storage.
In conclusion, the mitosis scaling approach can provide flexible and fine-grained scaling, effectively adapting to dynamic workload demands.

\subsection{Parallelism Compatibility}

\begin{figure}[!t]
    \centering    \includegraphics[width=0.95\linewidth]{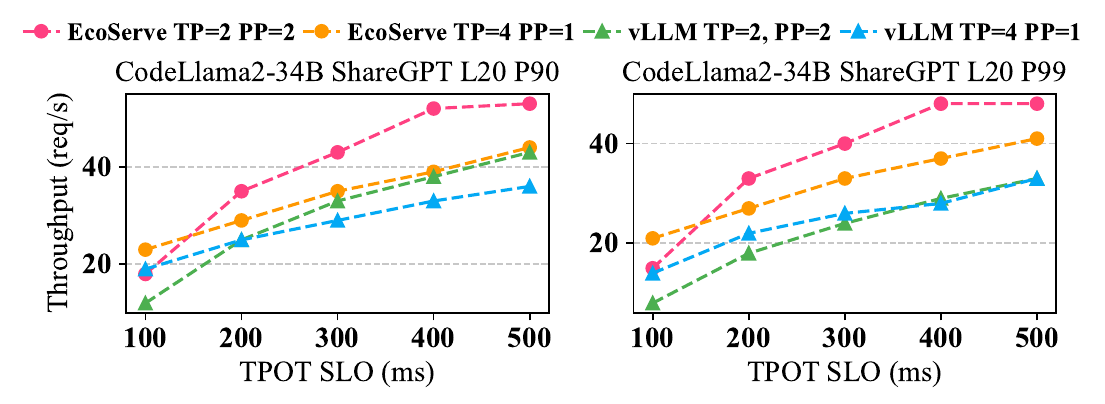}
    \caption{Pipeline parallel compatibility.}
    \label{fig:parallel_experiment}
\end{figure}

We further validate that the FuDG strategy is better suited for pipeline parallelism (PP) compared to the NoDG strategy.
We use CodeLlama2-34B, ShareGPT, and the L20 cluster.
As PP does not improve the latency of a single batch, a relaxed TPOT constraint is required.
Figure~\ref{fig:parallel_experiment} illustrates the throughput as the TPOT SLO increasing from 100ms to 500ms.
In this setup, CodeLlama2-34B is configured with TP=2, PP=2 and TP=4, PP=1.
It is evident that EcoServe, when utilizing PP, achieves better performance than its TP counterpart at lower TPOT SLOs, outperforming vLLM.
In other words, the intersection point occurs at a slower TPOT SLO and the throughput plateau achieved with PP is much higher than that of vLLM.

%% file: tex/Related.tex
\section{Related Work}

\noindent\textbf{Scheduling in LLM serving.}
Based on whether prefill and decode phases are disaggregated, existing LLM serving approaches can be categorized into the NoDG strategy~\cite{sglang2024github, vllm2024github, yu2022orca, agrawal2024taming} and the FuDG strategy~\cite{patel2024splitwise, distserve, qin2024mooncake}, which are most relevant to EcoServe.
Adrenaline~\cite{liang2025injecting} notices the load imbalance issue in FuDG and reschedules computation in prefill and decode instances.

Moreover, other studies address issues in specific inference scenarios.
Flexgen~\cite{sheng2023flexgen}, FastDecode~\cite{fastdecode2024} and Specinfer~\cite{miao2023specinfer} enable LLM inference with limited memory capacity by employing offloading strategies.
Loongserve~\cite{wu2024loongserve} and Infinite-llm~\cite{lin2024infinite} targets long-context inference and optimize parallel strategy and memory utilization respectively.
Moe-lightning \cite{cao2025moe}, Pre-gated MoE~\cite{hwang2024pre} and Lina~\cite{li2023accelerating} focus on MoE models and optimize resource utilization by employing expert popularity.
MegaScale-Infer~\cite{zhu2025megascaleinfer} targets ultra-large MoE model and it accelerates the decode phase by disaggregating the attention and FFN modules.
Liger~\cite{du2024liger} and NanoFlow~\cite{zhu2024nanoflow} carefully schedules and overlaps GPU kernels from different requests to improve efficiency.

\noindent\textbf{KV Cache Management.}
To reduce KV cache memory usage of standard MHA~\cite{vaswani2017attention}, MQA~\cite{shazeer1911fast} and GQA~\cite{ainslie2023gqa} share key and value projections across query heads.
PagedAttention~\cite{kwon2023efficient} and vAttention~\cite{prabhu2024vattention} reduces memory fragmentation by organizing the KV cache into fixed-size blocks.
To compress KV cache, H2O~\cite{zhang2023h2o}, Keyformer~\cite{adnan2024keyformer}, and Liu et al.~\cite{liu2024optimizing} find token similarity and removes redundant information.
Next, Shadowkv~\cite{sun2024shadowkv}, Prompt cache~\cite{gim2024prompt}, and Ragcache\cite{jin2024ragcache} further explore KV cache compression and offloading strategies in long-context scenarios.
AttentionStore~\cite{gao2024attentionstore} schedules KV cache across hierarchical storage tiers, while CacheBlend~\cite{yao2025cacheblend} introduces the pipelining loading with partial recomputation to use slower object stores.


%% file: tex/Discussion.tex
\begin{table}[!t]
\caption{Large-scale LLM Serving Strategy Comparison.}
\label{tab:large_comparision}
\centering
\resizebox{0.98\linewidth}{!}{
\begin{tabular}{ccccccc}
\hline
  & \textbf{Goodput} & \begin{tabular}[c]{@{}c@{}}\textbf{Cost} \\ \textbf{Effective}\end{tabular} & \begin{tabular}[c]{@{}c@{}}\textbf{Load} \\ \textbf{Balance}\end{tabular}  & 
  \begin{tabular}[c]{@{}c@{}}\textbf{Hardware} \\ \textbf{Cost}\end{tabular} & 
  \begin{tabular}[c]{@{}c@{}}\textbf{Parallelism} \\ \textbf{Compatibility}\end{tabular} &  
  \begin{tabular}[c]{@{}c@{}}\textbf{Engineering} \\ \textbf{Complexity}\end{tabular}\\ \hline \hline

NoDG   & \ding{51} & Good & Easy  & Low & Low & Low\\ 
FuDG  & \ding{51}\ding{51} & Poor & Hard & High & High & High \\ 
PaDG & \ding{51}\ding{51} & Excellent & Easy & Low & High & Low \\ \hline

\end{tabular}
}
\vspace{-10pt}
\end{table}
\section{Discussion}
Commercial success in LLM serving hinges on adopting cost-effective strategies on large-scale clusters, which \textbf{require a careful trade-off between throughput, SLO attainment, infrastructure cost, parallelism compatibility, and even engineering complexity}.
Table~\ref{tab:large_comparision} presents a comparison between the PaDG strategy and the existing NoDG and FuDG strategies.
In terms of goodput, the PaDG strategy is comparable to FuDG, while largely outperforming NoDG.
While FuDG is designed for tight SLOs and relies on high-performance interconnects, PaDG is optimized for cost-effective deployments.
When SLOs are relaxed or interconnects are limited, PaDG can outperform FuDG.

Beyond hardware, PaDG also reduces load imbalance and engineering complexity. Unlike FuDG, which scales across two instance types, both NoDG and PaDG scale at the granularity of individual instances, leading to simpler scaling and more balanced workloads.
In addition, the lack of cross-instance KV cache transmission in PaDG and NoDG significantly lowers system complexity.
From a parallelism compatibility perspective, PaDG offers further advantages:
its lower frequency of prefill-decode switching improves pipeline parallelism efficiency, while minimal data movement and reduced PCIe contention make it more suitable for tensor parallelism on systems without direct GPU interconnects.

\textbf{NoDG, PaDG, and FuDG each have their own advantageous scenarios, and LLM serving vividly demonstrates the art of trade-offs in system optimization.}
NoDG is well-suited for small models, such as 7B and 13B.
These models have lower computational demands, and their SLOs are easier to satisfy, making prefill-decode interference negligible.
Larger models, such as 30B, 70B, and 130B, benefit more from PaDG.
These models typically require parallel techniques to extend memory capacity and are still capable of meeting typical latency SLOs in a single instance.
In extreme scenarios, like ultra-large models or stringent SLOs, even minor interferences can significantly degrade these metrics, FuDG with advanced hardware becomes essential.
More aggressive strategies are also worth studying.
For example, MegaScale-Infer~\cite{zhu2025megascaleinfer} studies ultra-large MoE model and disaggregates the attention and FFN modules into different instances.
Moreover, these strategies incur incremental engineering costs, which also serve as a major factor.



%% file: tex/Conclusion.tex
\section{Conclusion}

This paper presents EcoServe, a cost-effective LLM serving system with a novel Partially Disaggregated (PaDG) Strategy.
The PaDG strategy leverages proactive intra and inter-instance scheduling to better balance TTFT and TPOT, significantly improving throughput on clusters with commodity interconnects.

